\documentclass[nofootinbib,notitlepage,superscriptaddress,10pt,aps,pra,twocolumn]{revtex4-2}

\usepackage{xcolor}
\usepackage[utf8x]{inputenc}
\usepackage{amsmath}
\usepackage{amssymb}
\usepackage{graphicx}
\usepackage{float}
\usepackage[normalem]{ulem}
\usepackage{epsf,epsfig}
\usepackage{psfrag}
\usepackage{color}
\usepackage{multirow}
\usepackage{diagbox}
\usepackage[hidelinks]{hyperref}
\usepackage{url}
\usepackage{xurl}
\usepackage{longtable}

\makeatletter
\newcommand\stroke[1]{\mathpalette\stroke@aux{#1}}
\def\stroke@aux#1#2{%
  \ooalign{%
    \hfil$#1^{\;\, \_\hspace{-0.05cm}\_}$\hfil\cr
    \hfil$#1#2$\hfil\cr
  }%
}
\makeatother

\begin{document}

\twocolumngrid 

\title{On testing in-vacuo dispersion with the most energetic neutrinos:\\
KM3-230213A case study}

\author{Giovanni Amelino-Camelia}
\affiliation{Dipartimento di Fisica Ettore Pancini, Universit\`a di Napoli “Federico II”, Complesso Univ. Monte S. Angelo, I-80126
Napoli, Italy}
\affiliation{Istituto Nazionale di Fisica Nucleare, Sezione di Napoli, Complesso Univ. Monte S. Angelo, I-80126 Napoli, Italy}
\author{Giacomo D'Amico}
\affiliation{Institut de Física d’Altes Energies (IFAE), The Barcelona Institute of Science and Technology (BIST), E-08193 Bellaterra, (Barcelona), Spain}
\author{Giuseppe Fabiano}
\affiliation{Physics Division, Lawrence Berkeley National Laboratory, Berkeley, CA 94720, USA}
\affiliation{Department of Physics, University of California, Berkeley, CA 94720, USA}
\affiliation{Centro Ricerche Enrico Fermi, I-00184 Rome, Italy}
\author{\\Domenico Frattulillo}
\affiliation{Istituto Nazionale di Fisica Nucleare, Sezione di Napoli, Complesso Univ. Monte S. Angelo, I-80126 Napoli, Italy}
\author{Giulia Gubitosi}
\affiliation{Dipartimento di Fisica Ettore Pancini, Universit\`a di Napoli “Federico II”, Complesso Univ. Monte S. Angelo, I-80126
Napoli, Italy}
\affiliation{Istituto Nazionale di Fisica Nucleare, Sezione di Napoli, Complesso Univ. Monte S. Angelo, I-80126 Napoli, Italy}
\author{Alessandro Moia}
\affiliation{Dipartimento di Fisica Ettore Pancini, Universit\`a di Napoli “Federico II”, Complesso Univ. Monte S. Angelo, I-80126
Napoli, Italy}
\author{Giacomo Rosati}
\affiliation{Dipartimento di Matematica, Università di Cagliari, via Ospedale 72, 09124 Cagliari, Italy}
\affiliation{Istituto Nazionale di Fisica Nucleare, Sezione di Cagliari,
Cittadella Universitaria, 09042 Monserrato, Italy}

\begin{abstract}

The phenomenology of in-vacuo dispersion, an effect such that quantum properties of spacetime slow down particles proportionally to their energies, has been a very active research area since the advent of the Fermi telescope. One of the assumptions made in this 15-year effort is that the phenomenology of in-vacuo dispersion has a particle-energy sweet spot: the energy of the particle should be large enough to render the analysis immune to source-intrinsic confounding effects but still small enough to facilitate the identification of the source of the particle. We use the gigantic energy of KM3-230213A as an opportunity to challenge this expectation. For a neutrino of a few hundred PeVs a transient source could have been observed at lower energies several years earlier, even assuming the characteristic scale of in-vacuo dispersion to be close to the Planck scale. We report that GRB090401B is in excellent directional agreement with KM3-230213A, and we discuss a strategy of in-vacuo-dispersion analysis suitable for estimating the significance of KM3-230213A as a GRB090401B-neutrino candidate. The $p$-value resulting from our analysis (0.015) is not small enough to warrant any excitement, but small enough to establish the point that a handful of such coincidences would be sufficient to meaningfully test in-vacuo dispersion. 
\end{abstract}

\maketitle

Motivated mainly by the fact that it could be caused by quantum properties of spacetime~\cite{grbgac, urrutia, COSTreview}, in-vacuo dispersion has been studied extensively over the last 15 years, trying to profit from the opportunity provided by modern telescopes like Fermi~\cite{FermiGBMLAT:2009nfe}. The relevant  analyses of astrophysical signals are tailored to find the imprint of the in-vacuo-dispersion excess contribution $\Delta t$ to the travel time of particles

\begin{equation}
\Delta t = D(z) \frac{E}{M_{QG}}   \, ,
\label{mainnewone}
\end{equation}

where $M_{QG}$ is the characteristic scale of in-vacuo dispersion, to be determined experimentally (but expected to be within 1 or 2 orders of magnitude~\cite{COSTreview} of 
the Planck scale $\sim 10^{16}$~TeV) and $D(z)$ is a function of the redshift $z$ of the source emitting the particle,

\[
D(z) = \int_0^z d\zeta \frac{(1+\zeta)}{H_0\sqrt{\Omega_\Lambda + (1+\zeta)^3 \Omega_m}}
\]

(as usual, $\Omega_\Lambda$, $H_0$ and $\Omega_m$ denote, respectively, the cosmological constant, the Hubble parameter and the matter fraction, for which we take the values given in Ref.~\cite{Planck:2018vyg}).

Hundreds of studies focused on the phenomenology of in-vacuo dispersion for photons~\cite{grbgac,COSTreview}, with some studies even reaching sensitivity of the order of the Planck scale (see, {\it e.g.}, Ref.~\cite{FermiGBMLAT:2009nfe}). 
The investigation of in-vacuo dispersion for neutrinos has been lagging far behind: the current bound on in-vacuo dispersion for neutrinos is still based on the historic observation of neutrinos from the SN1987a supernova, and only amounts to $M_{QG} \gtrsim 10^7 $ TeV~\cite{Ellis:2008fc} (which is $10^{-9}$ of the Planck scale). This is very unsatisfactory, especially in light of the fact that the quantum-spacetime literature provides  motivation for studying in-vacuo dispersion for photons and neutrinos separately. In particular, quantum-spacetime models based on Planck-scale discretization predict in-vacuo dispersion for neutrinos but not for photons~\cite{Alfaro:1999wd,Sahlmann:2002qj}, and this is consistent with results from models based on causal sets~\cite{Philpott:2010cm}. Moreover, within a popular effective-theory approach to quantum-spacetime  properties one describes in-vacuo dispersion for neutrinos in a way that is completely independent from that applicable to photons~\cite{Myers:2003fd}. 

\begin{figure*}[!htbp]
    \centering
    \includegraphics[scale=0.35]{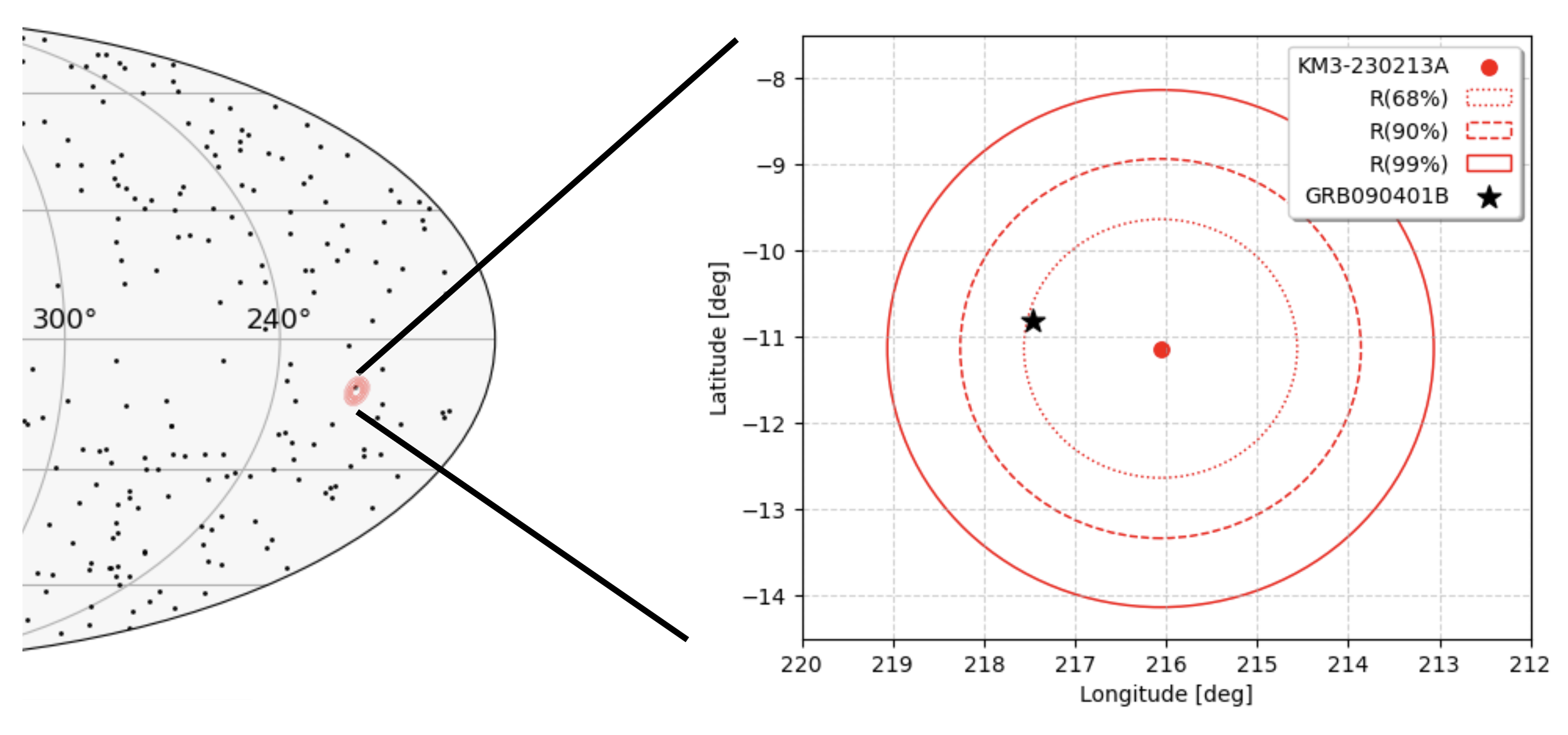}
    \caption{We here provide an assessment in galactic coordinates
    of the directional compatibility  between 090401B and 230213A. The dots on the left represent known-redshift GRBs (taken from the catalogue of Ref.~\cite{neutrini2025}) and the red circle on the left reflects the experimental information on the direction of 230213A~\cite{KM3NeT:2025npi} (the dot inside the red circle stands for 090401B). The right side quantifies visually the 090401B-230213A directional compatibility (see Appendix~\ref{Sdir}, and the corollary observations reported in Appendix~\ref{nozGRB}).}
    \label{fig1}
\end{figure*}

Recently, as the public availability of IceCube data increases~\cite{icecubedatarelease}, a few studies have been devoted to in-vacuo dispersion for neutrinos~\cite{NatureAstro, MaNeutriniFirst, Ellis:2018ogq, natastro2023, neutrini2025} (also see the earlier preliminary investigations~\cite{jacobpiran, gacGuettaPiran}). The prospect of observing neutrinos from GRBs (gamma-ray bursts) is particularly exciting from the in-vacuo-dispersion perspective, because of the fine time structure and typically large redshift of GRBs~\cite{grbgac, COSTreview}.

We here contemplate the possibility that the recently-announced ultra-high-energy neutrino 230213A (KM3-230213A~\cite{KM3NeT:2025npi}), with energy of $\sim\!\!\!220$ PeV, might be a GRB neutrino. A priori, this would seem a realistic scenario. Indeed, in light of the bounds set by IceCube on the flux of cosmogenic neutrinos, a cosmogenic origin for 230213A is regarded as very unlikely~\cite{KM3NeT:2025npi,kahunaNOcosmogenic}. And blazars are not expected to have enough fire power for the observation of such a high-energy neutrino~\cite{kahunaNOblazar}. A GRB origin for 230213A would instead make observational and theoretical sense (see, {\it e.g.}, Ref.~\cite{kahunagrb}), but one finds no GRB in good directional and temporal agreement with 230213A. However, the in-vacuo-dispersion hypothesis leads us to question the standard requirement that a ultra-high-energy GRB neutrino should be observed in close temporal coincidence with the much-lower-energy electromagnetic GRB signal. In fact, in the presence of in-vacuo dispersion, a 220-PeV GRB neutrino could very well be delayed by several years, even assuming that $M_{QG}$ is rather close to the Planck scale.

We observe that, as shown in Fig.~\ref{fig1}, the estimated direction of the 230213A neutrino is consistent with the direction of the GRB 090401B (GRB090401B~\cite{GRB090401Bgcn}), with measured redshift 3.1 (see Appendix~\ref{Sdir}). Moreover, the 14-year separation between the two observations would be unsurprising from an in-vacuo-dispersion standpoint, considering both the remarkably high energy of 230213A and the relatively high redshift of 090401B.
Because of how crowded our Universe is, the burden of proof for arguing a connection between observations separated by 14 years must be very onerous, and indeed we here find (see later) that with presently-available data (and data uncertainties) the case for attributing 230213A to 090401B is not very strong. But what would be needed to make such kind of ``GRB-neutrino candidate" statistically significant? Is it at all possible? Would it make a big difference if the direction of 230213A had been determined more sharply? What if at some point we had a few such GRB-neutrino candidates, with large time separation between the observation of the GRB and the observation of the ultra-high-energy neutrino?

For the Fermi-telescope phenomenology of in-vacuo dispersion with GRB photons of multi-GeV energies the challenge is complementary: since a Planck-scale estimate of the magnitude of in-vacuo dispersion is in that case of no more than $\sim\!\!100$ seconds, the identification of the source is straightforward, but then the effect is small enough for its study to  be weakened by some known (but poorly understood) source-intrinsic effects which reduce the sensitivity reach of the analysis~\cite{COSTreview, FermiGBMLAT:2009nfe}. For GRB neutrinos the ``sweet spot" could be neutrino energies between 100 and 500~TeV~\cite{NatureAstro}, for which the magnitude of the effect should be big enough to render source-intrinsic effects irrelevant, but still small enough to give realistic chances of source identification.  

We here assess the statistical significance of the 090401B-230213A association by relying on a suitable adaptation of the approach developed in Refs.~\cite{NatureAstro, natastro2023, neutrini2025}, an adaptation which is made necessary by the challenges posed by the huge energy (and huge energy uncertainty) of 230213A. The key aspect of our strategy of analysis is that one should quantify in terms of some statistic $\mathcal{S}_{dir}$ the level of directional agreement between the GRB and the neutrino and quantify in terms of some other statistic $\mathcal{S}_E$ the level of agreement between the experimental information on the energy of the neutrino and the range of energies that in-vacuo dispersion would predict for such a delayed neutrino observation assuming it has been emitted at the GRB redshift. One can then estimate how likely it would be to produce accidentally, if the neutrino was unrelated to known-redshift GRBs, values of $\mathcal{S}_{dir}$ and $\mathcal{S}_E$ as high as those of the studied GRB-neutrino pair.

For what concerns $\mathcal{S}_{dir}$, a natural candidate has already been considered in previous multimessenger studies~\cite{neutrini2025, LIGOScientific:2017zic}: the statistic $\mathcal{S}_{dir}=\int P_{\nu}(\Omega)P_{GRB}(\Omega)d\Omega$, where $P_{\nu}(\Omega)$ and $P_{GRB}(\Omega)$ are the angular distributions of the neutrino and the GRB, respectively, is a good measure of directional compatibility. As visually clear from Fig.~\ref{fig1}, we find that the  090401B-230213A  association  has indeed a rather large $\mathcal{S}_{dir}$, equal to 194 (see Appendix~\ref{Sdir}, and the corollary observations reported in Appendix~\ref{nozGRB}).

\begin{figure}[H]
    \centering
    \includegraphics[scale=0.9]{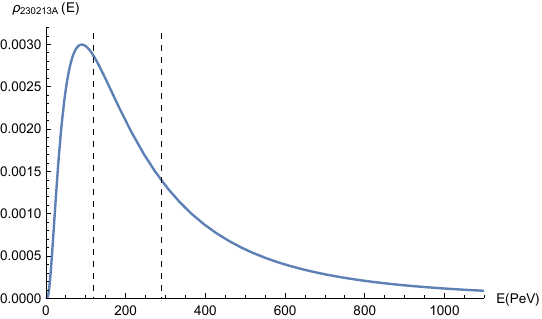}
    \caption{Here the two vertical lines mark the extrema ($119$~PeV and $289$~PeV) of the range of inferred 230213A energy,  obtained using Eq.~\eqref{mainnewone} and the range of values for $M_{QG}$ motivated in Ref.~\cite{natastro2023}. $\rho_{230213A}(E)$, which we used to compute our statistic $\mathcal{S}_E$, is a log-normal distribution describing the energy of 230213A, with the properties that  its median is $220$~PeV, its integral between $110$~PeV and $790$~PeV is $0.68$ and its integral between $72$~PeV and $2600$~PeV is $0.9$~\cite{KM3NeT:2025npi}. }
    \label{fig2}
\end{figure}

There are no previous proposals for the statistic $\mathcal{S}_E$. The $\mathcal{S}_E$ we here propose takes as starting points the redshift $z$ of the GRB, equal to $z_{\mathrm{090401B}} =3.1$ in the case of 090401B, and the $\Delta t$ between the neutrino observation time and the GRB observation time, which in the case of the 090401B-230213A pair is $\Delta t_{\mathrm{090401B}} = 4.38\cdot  10^8$~s. In light of the in-vacuo dispersion of Eq.~\eqref{mainnewone}, the measured values of $\Delta t$ and $z$ allow us to convert any chosen value of $M_{QG}$ into an in-vacuo-dispersion-inferred value of the neutrino energy. The analysis reported in Ref.~\cite{natastro2023} led to estimating that $M_{QG}$ could be in the range $[3.97\cdot 10^{14},9.60\cdot 10^{14}]$~TeV, using IceCube neutrinos with energy between $60$ and $500$~TeV and making rather crude assumptions on the redshift of some GRBs whose redshift was not measured. Then in Ref.~\cite{neutrini2025} this same range of values for $M_{QG}$ was the core ingredient for a rather encouraging analysis (on which we find appropriate to comment in more detail later in this manuscript) restricted to GRBs of known redshift\footnote{In Refs.~\cite{natastro2023} and~\cite{neutrini2025} the in-vacuo-dispersion scale $M_{QG}$ was equivalently characterized in terms of the Planck scale $M_{Planck}$ and the dimensionless parameter $\eta= M_{Planck}/M_{QG}$.}. For the investigation of the 090401B-230213A pair we take once again as reference the range $[3.97\cdot 10^{14},9.60\cdot 10^{14}]$~TeV for $M_{QG}$. For any given GRB-neutrino pair, using its measured values of $\Delta t$ and $z$, this range of values for $M_{QG}$ is converted, on the basis of Eq.~\eqref{mainnewone}, into a range $[E_{min},E_{max}]$ for the inferred neutrino energy $E$. In the case of the 090401B-230213A pair  we find $E_{min} = 119$~PeV and $E_{max}=289$~PeV. A good $\mathcal{S}_E$ statistic should efficaciously characterize the quantitative agreement between this range of (in-vacuo-dispersion-)inferred energies for the neutrino and the experimentally measured range of energies for the neutrino, which in general will take the form of some energy distribution $\rho_\nu(E)$. Our $\mathcal{S}_E$ statistic is given by

\[
\mathcal{S}_E = \int_{E_{min}}^{E_{max}}  \frac{\rho_{\nu}(E)}{E_{\tiny{{max}}}-E_{\tiny{{min}}}}\, dE \, ,
\]

which in the case of the 090401B-230213A pair takes the value $\mathcal{S}_E = 0.00209$~PeV$^{-1}$, reflecting the fact that, as shown in Fig.~\ref{fig2}, a significant portion of $\rho_{\mathrm{230213A}}(E)$ falls within the range $[119,289]$~PeV.

In order to assess the significance of our findings of $\mathcal{S}_{dir}$ ($=194$) and $\mathcal{S}_E$ ($=0.00209$~PeV$^{-1}$), we produce simulated GRB data and use them to establish how frequently values of $\mathcal{S}_{dir}$ and $\mathcal{S}_E$ as high as those of the 090401B-230213A pair would be found accidentally if 230213A was actually unrelated to known-redshift GRBs. To produce our simulated GRB data we use a rather standard approach (see, {\it e.g.}, Ref.~\cite{LIGOScientific:2017zic}) as adapted to known-redshift GRBs in Ref.~\cite{neutrini2025}: each instance of simulated GRB data is obtained from the known-redshift GRB catalogue of Ref.~\cite{neutrini2025} by acting on the GRB directions with a random permutation and a random rotation around the Galactic axis (see Ref.~\cite{neutrini2025} for further details). Generating $10^5$ such instances of simulated data we found that values of $\mathcal{S}_{dir}$ and $\mathcal{S}_E$ as high as (or higher than) those of the 090401B-230213A pair occur accidentally only $1.5\%$ of the times, yielding a $p$-value of 0.015 (which corresponds to a $2.4\sigma$ significance in Gaussian statistics).

Having found that the association
090401B-230213A
is intriguing but far from conclusive, we
can explore possible avenues for a similar GRB-neutrino candidate to be more significant. One realistic avenue is for the uncertainty in the direction of the neutrino to be smaller.
It is realistic even for 230213A itself since its present directional uncertainty of $1.5^{\circ}$ is dominated by systematics for which the KM3NeT collaboration is planning improvements~\cite{KM3NeT:2025npi}. As shown in Appendix~\ref{Uncertainty},  for a GRB-neutrino pair like 090401B-230213A,
but assuming a
 directional uncertainty of 
$0.2^{\circ}$ for 230213A,
an analysis such as ours should achieve a $p$-value of $\sim 0.0002$.

It is difficult to estimate how frequently we might have observations of neutrinos with energy comparable 
to 230213A. Based on results reported by IceCube~\cite{icecubegen2}
one might expect 
this frequency to be rather low. It is nonetheless worth stressing that if one day we could have a handful of such GRB-neutrino candidates, with individual significance of about $1\%$, but different (though comparable) energies,
then 
the data could be also used to 
test~\cite{NatureAstro, natastro2023, neutrini2025}
the energy-dependence predicted 
by Eq.~\eqref{mainnewone}, strengthening the significance of the analysis.
In this respect there is a particularly amusing scenario that can be contemplated: the
observation within a few of years of a neutrino with energy  $20\%$ or $30\%$ higher than 230213A
and direction once again compatible with 090401B.

\begin{figure}[H]
\centering
\includegraphics[scale=0.29]{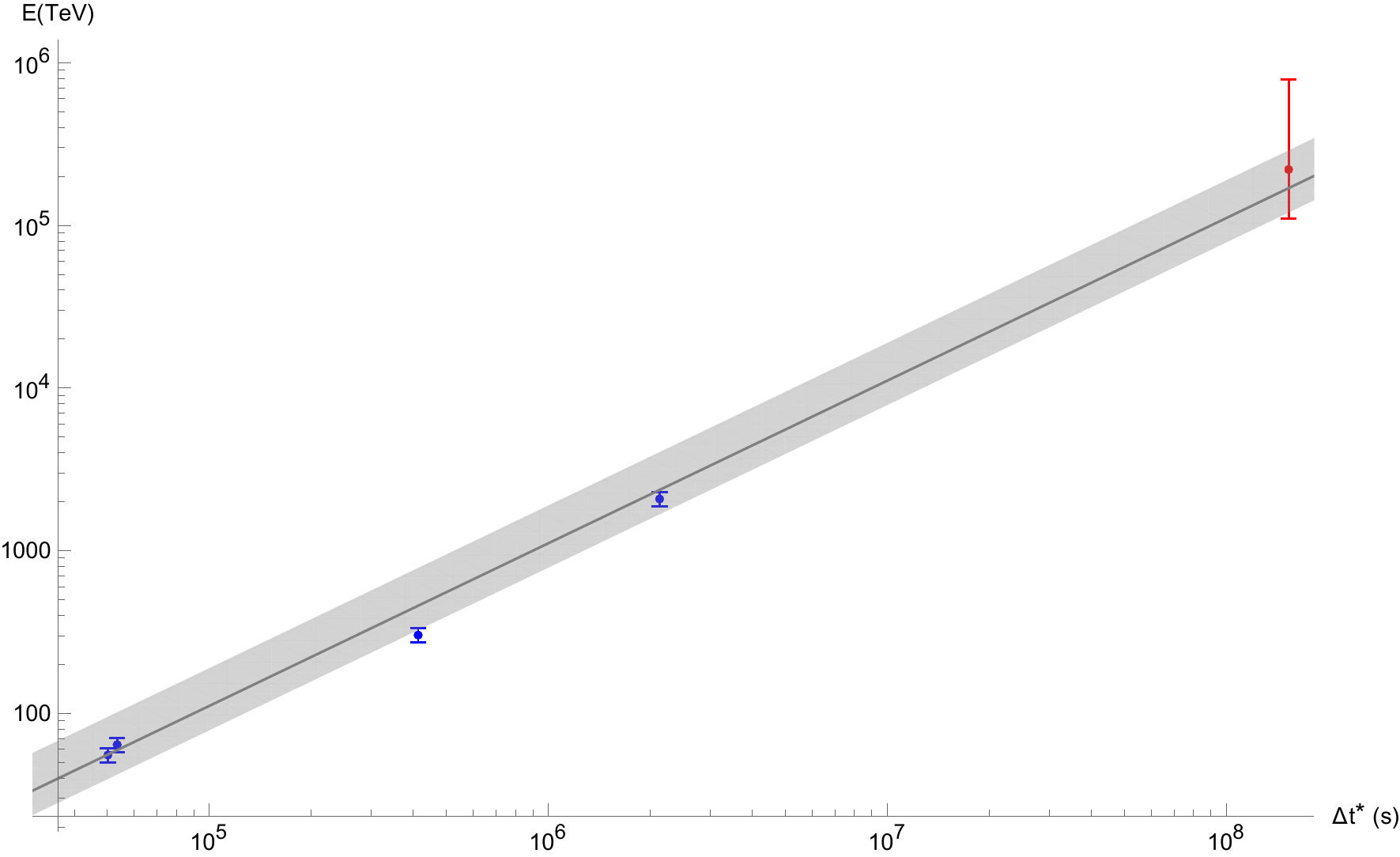}
    \caption{
Here  $\Delta t^*$ is $\Delta t D(1)/D(z)$
(with $D(1)$ introduced only for the convenience of having a  $\Delta t^*$ with dimensions of time). According to Eq.~\eqref{mainnewone}, for GRB neutrinos there should be a linear relationship between $E$ and $\Delta t^*$ (when $\Delta t^*$ is computed using the redshift $z$ of the GRB and the  $\Delta t$ between neutrino and GRB). The gray band and the gray line reflect properties of the search window first motivated in Ref.~\cite{natastro2023}. The highest-energy (red) data point corresponds to the 090401B-230213A pair, while the other four data points represent the four GRB-neutrino candidates found in Ref.~\cite{neutrini2025}.}
    \label{fig3}
\end{figure}

Even if 230213A remained unique, and if somehow its directional uncertainty was not improved through better control of KM3NeT systematics,
the 
090401B-230213A pair could still contribute its $ 1.5 \%$ significance to studies of in-vacuo dispersion also involving neutrinos of lower energies.
The content of Fig.~\ref{fig3} refers to  
the 090401B-230213A pair and to four other GRB-neutrino candidates found in Ref.~\cite{neutrini2025}. The study reported 
in Ref.~\cite{neutrini2025} adopted the same $M_{QG}$ search window here adopted 
($3.97\cdot 10^{14}\, \text{TeV} < M_{QG} < 9.60\cdot 10^{14}\, \text{TeV}$, which is reflected in the gray band of Fig.~\ref{fig3})
and the four GRB-neutrino candidates of 
Ref.~\cite{neutrini2025} had an overall significance, from the in-vacuo-dispersion perspective, of $0.6\%$  ($2.8 \sigma$
significance in Gaussian statistics).

The content of Fig.~\ref{fig3} shows qualitatively a path toward a role for the 090401B-230213A pair in studies of in-vacuo dispersion also involving lower-energy neutrinos. In particular, there is a rather strong consistency,
from the in-vacuo-dispersion perspective,
between the 090401B-230213A pair
and the four GRB-neutrino candidates highlighted in
Ref.~\cite{neutrini2025}. Evidently, this consistency could  motivate an effort aimed at establishing the combined statistical significance for in-vacuo dispersion of all five GRB-neutrino candidates appearing in Fig.~\ref{fig3}.
We postpone that assessment to a future study, since we have not yet been able to find a satisfactory way to combine the approaches introduced here and in Ref.~\cite{neutrini2025}: the strategy of analysis here introduced is tailored for the task of assessing the significance of a single GRB-neutrino pair, and is robust enough to handle a large uncertainty in the neutrino energy, while the strategy of analysis of Ref.~\cite{neutrini2025} relies strongly on the energy dependence 
predicted by Eq.~\eqref{mainnewone}, and therefore requires neutrinos whose energy is sharply determined experimentally.

\noindent
We devote one final remark to the possibility, also much studied in the quantum-gravity literature, that in-vacuo dispersion might be quadratic rather than linear \cite{COSTreview}:
$\Delta t = D'(z) E^{\,^2}/{M'^{\,^2}_{QG}}$. The directional agreement of the 090401B-230213A pair is still assessed in the same way from the perspective of quadratic in-vacuo-dispersion, and with a single GRB-neutrino candidate one evidently cannot probe the energy dependence of the effect.
If however one attributed any relevance to the consistency between 
090401B-230213A 
and the four other GRB-neutrino pairs here discussed in relation to Fig.~\ref{fig3}, then only linear dispersion would be viable.

\begin{acknowledgments}
G.A.-C. and G.G. are grateful for financial support by the Programme STAR Plus, funded by Federico II University and Compagnia di San Paolo,
 and by the MIUR, PRIN 2017
grant 20179ZF5KS. G.D.’s work on this project was supported by the Beatriu de Pin\'{o}s programme (BP 2023). G.F.'s work on this project was supported by ``The Foundation Blanceflor".
This work also falls within the
scopes of the  COST Action CA23130 ``Bridging high and low energies in search of quantum gravity” and the COST Action CA18108 ``Quantum Gravity phenomenology in the Multi-Messenger era".
\end{acknowledgments}

\appendix

\section{GRB090401B, KM3-230213A and $\mathcal{S}_{dir}$}\label{Sdir}

We here collect some information on 090401B and 230213A, focusing on what is relevant for the evaluation of $\mathcal{S}_{dir}$
for the pair composed by 090401B and 230213A.

090401B is a long GRB ($\mathrm{T90}=183$~s), with measured redshift 3.1~\cite{refGRB090401Bredshift}, whose J2000 coordinates are $\mathrm{RA}=95.088^{\circ}$, $\mathrm{decl}=-8.972^{\circ}$ with an angular uncertainty of $0.000065^{\circ}$
\cite{GRB090401Bgcn}.

The J2000 coordinates of 230213A are $\mathrm{RA}=94.3^{\circ}$, $\mathrm{decl}=-7.8^{\circ}$ with containment radius $\mathrm{R}(68\%)=1.5^{\circ}$~\cite{KM3NeT:2025npi}. (The available experimental information on the energy of 230213A is summarized in the main text.)

To characterize the level of directional agreement between 090401B and 230213A we use the statistic 
$\mathcal{S}_{dir}$ given by $\int P_{\mathrm{230213A}}(\Omega)P_{\mathrm{090401B}}(\Omega)d\Omega$, where $P_{\mathrm{230213A}}(\Omega)$ and $P_{\mathrm{090401B}}(\Omega)$ are the angular distributions
of 230213A and 090401B, respectively.
We assume that 
$P_{\mathrm{230213A}}(\Omega)$ and $P_{\mathrm{090401B}}(\Omega)$ are Gaussian distributions characterized by the available experimental information on direction and directional uncertainty for 
230213A and 090401B. Actually, because of the large difference in directional uncertainties, the angular distribution of 090401B can be approximated with a $\delta$-function when computing $\mathcal{S}_{dir}$. We find $\mathcal{S}_{dir}=194$.  The associated (purely directional) statistical significance can be easily determined looking at how frequently values of $S_{dir}$ of 194 or higher occur accidentally in our simulated data (see the main text), and we find that this probability is of $5.5\%$.
Requiring that also $S_{E} \geq 0.00209$ PeV$^{-1}$ (the value of $S_{E}$ obtained for the 090401B-230213A pair), one obtains the $p$-value 0.015 quoted in the main text, which quantifies the overall significance of the 090401B-230213A association for the in-vacuo-dispersion scenario studied here.

\section{Aside on GRBs of unknown  redshift}\label{nozGRB}

The results for the significance of the 090401B-230213A pair reported in the main text provide quantitative support for our choice to immediately focus our in-vacuo-dispersion investigations on known-redshift GRBs observed several years earlier than 230213A. A known-redshift GRB like 090401B effectively provides a sharp spacetime point probing the in-vacuo dispersion of 230213A. The spacetime localization of 090401B is only limited by the minute uncertainty of its direction, the minute uncertainty of its redshift and by its overall duration in time, all of which are completely negligible, since, within an in-vacuo-dispersion analysis, the large energy uncertainty of 230213A translates into a large uncertainty in the spacetime point  of its emission.

This can perhaps be even better appreciated by contemplating the case of a neutrino with energy $\sim\!\!100$~PeV emitted by a persistent source characterized by intermittent flaring, like a blazar~\cite{IceCubeTXS}. In that case direction and redshift would be sharp but one would not have the sharp identification of  $\Delta t$ needed for the in-vacuo-dispersion analysis: if there are more candidate values of $\Delta t$ (one for each flare) separated by, say, several months, it will be clearly impossible to conclusively associate, through in-vacuo dispersion, a certain flare to a neutrino observed years later with a large energy uncertainty. Of course, in-vacuo dispersion does not exclude in any way that a blazar could produce a neutrino with energy $\sim\!\!100$~PeV. In particular, even assuming in-vacuo  dispersion, it is possible that 230213A was actually emitted by a blazar, but, in that case, an analysis such as ours could never identify a candidate source within the in-vacuo-dispersion scenario considered here. GRBs (or other non-repeating, ``one-shot" transients) would be our only chance to localize rather sharply the spacetime region of emission of a neutrino with energy $\sim\!\!100$~PeV affected by the kind of in-vacuo dispersion investigated here.

In light of this key role played by GRBs for in-vacuo-dispersion studies of the most energetic neutrinos, we found appropriate to perform a corollary analysis concerning GRBs whose redshift is unknown. It happens occasionally that the redshift of a GRB is determined several years after its observation  (see, {\it e.g.}, Refs. \cite{Kruhler:2012tz,Chrimes:2018ptj}). This encouraged us to  search for GRBs whose redshift is currently unknown that might be directionally compatible with 230213A. In principle, if this search  found a GRB whose direction was in particularly good agreement with the direction of 230213A, there would be further motivation, from the in-vacuo-dispersion perspective, to investigate the redshift of that GRB (though of course in most cases it would not be possible). However, we interestingly find no such case: 090401B is the GRB with the best directional agreement with 230213A, even including GRBs of unknown redshift.

For this corollary analysis we relied on the catalogues \cite{swiftGRBs} and \cite{icecubeGRBs}, providing directional information on 3159 GRBs whose redshift is unknown (while there are only 652 known-redshift GRBs in the catalogue of Ref.~\cite{neutrini2025}).
For each of these GRBs we quantified the directional agreement with 230213A using the statistic $\mathcal{S}_{dir}$, finding that, for all of them, the value of $\mathcal{S}_{dir}$ is significantly smaller than 194 (which is the value of $\mathcal{S}_{dir}$ for the 090401B-230213A pair).

\begin{figure}[h!]
    \centering
    \includegraphics[scale=0.35]{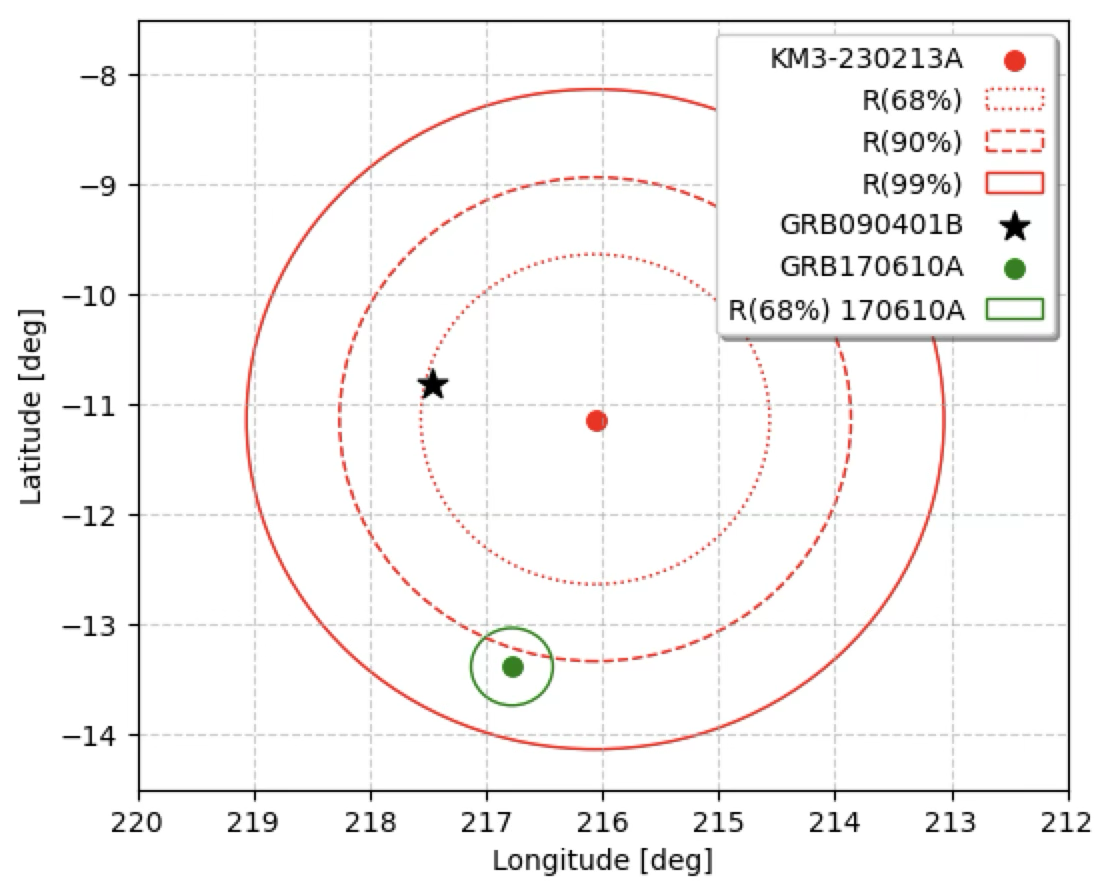}
    \caption{We here quantify visually the 090401B-230213A directional compatibility (already shown in Fig.~\ref{fig1}), corresponding to $\mathcal{S}_{dir}=194$, and the directional compatibility between GRB170610A and 230213A, corresponding to $S_{dir}=36$.}
    \label{fig4}
\end{figure}

In Fig.~\ref{fig4} we provide a visual comparison of the directional agreement for the 090401B-230213A pair and for the pair composed by GRB170610A and 230213A. GRB170610A is a fast transient X-ray source~\cite{icecubeGRBs} with J2000 coordinates $\mathrm{RA}=92.6^{\circ}$ $\mathrm{decl}=-9.4^{\circ}$ and an angular uncertainty of $0.3^{\circ}$, and is the GRB of unknown redshift with the best directional agreement with 230213A, 
agreement characterized by 
$\mathcal{S}_{dir}=36$. The value of $S_E$ cannot be computed since the redshift of  GRB170610A is unknown, but in any case already the directional compatibility between GRB170610A
and the 230213A neutrino is not noteworthy:
using simulated data we find that there is a probability of $61 \%$ of having accidentally a GRB with $\mathcal{S}_{dir}$
of 36 or higher.

After we announced a preprint version of this manuscript, a partly-related study, searching for possible GRB counterparts of the 230213A neutrino under the assumption of in-vacuo dispersion, was reported in Ref.~\cite{Wang:2025lgn}. Rather than quantifying the significance of the directional agreement of counterpart candidates with 230213A, as we did in our corollary analysis, the authors opted for just listing all GRBs satisfying certain general criteria of potential directional compatibility with 230213A (including of course 090401B and the above-mentioned GRB170610A).

 Our analysis shows that, based on presently available directional evidence, 090401B is definitely the most compelling candidate GRB counterpart of 230213A. Just to make an example, the significance of GRB920711A* (one of the GRBs highlighted in Ref.~\cite{Wang:2025lgn}) as a candidate GRB counterpart of 230213A, quantified by our directional statistic $\mathcal{S}_{dir}$, amounts to $\mathcal{S}_{dir}=7$, much lower than both $\mathcal{S}_{dir}=194$ (corresponding to 090401B) and $\mathcal{S}_{dir}=36$ (corresponding to GRB170610A). This is due to the huge directional uncertainty of GRB920711A* ($13^{\circ}$), which considerably weakens the implications of its estimated direction being relatively close (within $0.6^{\circ}$) to that of 230213A (the probability that
two directions coincide cannot be high when one of the two directions is very uncertain). Using simulated data
we find that there is a probability of $92 \%$ of having accidentally a GRB with $\mathcal{S}_{dir}$
of 7 or higher. Nonetheless, the approach of Ref.~\cite{Wang:2025lgn} is usefully complementary to ours: future findings may lead to a revision of our conclusions, and the GRB list of Ref.~\cite{Wang:2025lgn} would be the natural starting point of any such reassessment. In particular, when the KM3NeT collaboration recalibrates the detector~\cite{KM3NeT:2025npi}, thereby reducing the large systematic component of the directional uncertainty of the 230213A neutrino, some of the other GRBs highlighted in Ref.~\cite{Wang:2025lgn} may become more significant.

\vspace{0.4cm}

\section{On the implications of a hypothetical sharper determination of the direction of 230213A}\label{Uncertainty}

We here devote a small investigation to quantify the observation that in our assessment of the significance of the 090401B-230213A pair the uncertainty on the direction of 230213A is a key limitation.

For this purpose we replace the real directional uncertainty of 230213A, which is  $1.5^{\circ}$, with $0.2^{\circ}$. The real 090401B has angular distance of $1.4^{\circ}$ from the direction of 230213A, but for the purpose of this investigation we rescale this angular distance to $ 0.187^{\circ}$. The resulting value for our directional statistic would then be $\mathcal{S}_{dir}=10993$. With these changes (and leaving all other aspects of the analysis unchanged)
we redid the whole analysis, finding in our hypothetical scenario a $p$-value of about 0.00024.

\end{document}